\documentclass[aps,pra,twocolumn,preprintnumbers,amsmath,amssymb,superscriptaddress]{revtex4-1}
\usepackage{graphicx}
\usepackage[caption=false]{subfig}
\usepackage{epsfig}
\usepackage{dcolumn}
\usepackage{bm}
\usepackage{ulem}
\usepackage{color}
\usepackage{float}
\usepackage[colorlinks]{hyperref}
\usepackage{verbatim}
\hypersetup{citecolor = blue}

\def\bra#1{\left\langle#1\right|}
\def\ket#1{\left|#1\right\rangle}

\def\abs#1{\left|#1\right|}

\def\Re{{\rm Re}}
\def\Im{{\rm Im}}

\def\be{\begin{equation}}       \def\ee{\end{equation}}
\def\bea{\begin{eqnarray}}      \def\eea{\end{eqnarray}}
\def\ba{\begin{array}}
\def\ea{\end{array}}
\def\bnum{\begin{enumerate} }
\def\enum{\end{enumerate}}

\def\=>{\Rightarrow}
\def\>{\rightarrow}

\def\eye2{Fathbb{I}}

\def\Eq#1{Eq.~(\ref{#1})}
\def\Fig#1{Fig.~\ref{#1}}

\renewcommand{\>}{\rangle}
\renewcommand{\Im}{{\rm Im}}
\renewcommand{\Re}{{\rm Re}}
\newcommand{\p}{\partial}

\newcommand{\eq}[2]{
	\begin{equation}
	#1 \label{#2}
	\end{equation}
}

\renewcommand{\rm}[1]{\mathrm{#1}}

\newcommand{\vect}[1]{\boldsymbol{#1}}

\usepackage{listings}
\usepackage{color}
\definecolor{lightgray}{gray}{1}

\newcommand{\inlinecode}[2]{\colorbox{lightgray}{\lstinline[language=#1]$#2$}}

\newtheorem{theorem}{Theorem}
\newtheorem{corollary}{Corollary}[theorem]
\newtheorem{lemma}{Lemma}

\begin{document}

\title{Automatic Differentiable Monte Carlo: Theory and Application}

\author{Shi-Xin Zhang}
\thanks{These two authors contributed equally to this work}
\affiliation{Institute for Advanced Study, Tsinghua University, Beijing 100084, China}
\author{Zhou-Quan Wan}
\thanks{These two authors contributed equally to this work}
\affiliation{Institute for Advanced Study, Tsinghua University, Beijing 100084, China}
\author{Hong Yao}
\email{yaohong@tsinghua.edu.cn}
\affiliation{Institute for Advanced Study, Tsinghua University, Beijing 100084, China}
\affiliation{Department of Physics, Stanford University, Stanford, California 94305, USA}

\begin{abstract}
	Differentiable programming has emerged as a key programming paradigm empowering rapid developments of deep learning while its applications to important computational methods such as Monte Carlo remain largely unexplored. Here we present the general theory enabling infinite-order automatic differentiation on expectations computed by Monte Carlo with \textit{unnormalized} probability distributions, which we call ``automatic differentiable Monte Carlo'' (ADMC). By implementing ADMC algorithms on computational graphs,  one can also leverage state-of-the-art machine learning frameworks and techniques to traditional Monte Carlo applications in statistics and physics. We illustrate the versatility of ADMC by showing some applications: fast search of phase transitions and accurately finding ground states of interacting many-body models in two dimensions. ADMC paves a promising way to innovate Monte Carlo in various aspects to achieve higher accuracy and efficiency, e.g.  easing or solving the sign problem of quantum many-body models through ADMC.
\end{abstract}

\date{\today}
\maketitle

\section{Introduction}
Differentiation is a broadly important concept and a widely useful method in subjects such as mathematics and physics. 
Automatic differentiation (AD) evaluates derivatives of any function specified by computer programs \cite{Bartholomew-Biggs2000a, atlm2018} by propagating derivatives of primitive operations via chain rules. 
It is different from conventional symbolic differentiations by totally avoiding complicated analytic expressions of derivatives and is advantageous to numerical differentiations by totally eliminating discretization errors.
Besides, AD is particularly successful in calculating higher-order derivatives and computing gradients with respect to large number of variables as the case for gradient-based optimization algorithms.
Emerging as a new programming paradigm, AD is now extensively utilized in machine learning. Being one of the most important infrastructure for machine learning, it enables massive exploration on neural networks structures.

The great application potential of AD in fields beyond machine learning started to emerge.  Specifically, AD has been applied to certain areas of computational physics; for instance its interplay with tensor networks were investigated very recently \cite{Liao2019, Torlai2019a, Hubig2019, Wan2019}. One may naturally ask whether AD can be leveraged to Monte Carlo (MC) methods, another big family of computational algorithms. Initial investigations on encoding MC methods into AD framework \cite{Mohamed2019a} all assumed normalized probability distributions for Monte Carlo sampling. However, for nearly all interesting problems the normalization factor is not known {\it a priori} and the probability distribution is usually unnormalized. Fortunately, knowing the ratio of probabilities between different configurations is sufficient to perform MC simulations in Metropolis-Hasting algorithm \cite{Hastings1970}; Monte Carlo with unnormalized probability distribution is now a widely employed numerical approach in statistics and physics. Consequently, it is highly desired to integrate AD into generic Monte Carlo to achieve high accuracy and efficiency in various MC applications such as solving interacting many-body models in physics.

In this paper, we fill in the gap by proposing the general theory enabling infinite-order automatic differentiation on expectations computed by Monte Carlo with \textit{unnormalized} probability distributions, which we call ``automatic differentiable Monte Carlo'' (ADMC).  
Specifically, ADMC employs the method of AD to compute gradients of MC expectations, which is a key quantity used in statistics and machine learning \cite{Mohamed2019a}, without {\it a priori} knowledge of normalization factor or partition function which is the case for nearly all application scenarios in Markov chain Monte Carlo (MCMC). 
As MC gradient problem lies at the core of probabilistic programming \cite{Meent2018} and plays a central role in various fields including optimization, variational inference, reinforcement learning, and variational Monte Carlo (VMC), ADMC can be employed to a wide range of MC applications to achieve high accuracy and efficiency.

ADMC not only works with gradient of expectations in MC with unnormalized distributions but also holds true for higher-order derivatives of MC expectations. 
In contrast, MC estimation of higher order derivatives were rarely considered \cite{Foerster2018b}.
In addition, ADMC can be embedded in general stochastic computational graph \cite{Schulman2015b} framework seamlessly and play a critical role at the interplay between differentiable programming and probabilistic programming.
By introducing ADMC, we can build Monte Carlo applications in the state-of-the-art machine learning infrastructure to achieve high accuracy and efficiency in addressing questions such as fast search of phase transitions and ground states of interacting quantum models. For models we studied by ADMC, comparable or higher accuracy has been obtained comparing with previous methods such as RBM and tensor network. Moreover, ADMC paves a promising way to innovate Monte Carlo in various aspects, e.g. easing or solving the sign problem \cite{Loh1990, Wu2005, Berg2012, Huffman2014, Li2015b, Wang2015, Li2016a, Wei2016, Broecker2017, Li2018} of quantum Monte Carlo (QMC) \cite{Blankenbecler1981, CEPERLEY1986, Sandvik1998, Prokofev1998, Gull2011}.

The organization of this paper is as follows. In Sec.~\ref{sec:background}, we review important background knowledge required to understand the general theory of ADMC and its applications, including automatic differentiation, estimation on Monte Carlo gradients, and variational Monte Carlo methods. 
In Sec.~\ref{sec:theory}, we elaborate our theory towards ADMC, including detach function, ADMC estimator for normalized and unnormalized probability distributions as well as general theory on Fisher information matrix (FIM) with unnormalized probabilities , and the general theory for the AD-aware version of VMC. In Sec.~\ref{sec:application}, we present two explicit ADMC applications in physics: fast search of phase transitions and critical temperature in 2D Ising model; and end-to-end general-purpose ADVMC algorithms and accurately finding the ground state of the 2D quantum spin-1/2 Heisenberg model. We demonstrate how to leverage the power of state-of-the-art machine learning to ADMC algorithms in particular. In Sec.~\ref{sec:discussions}, we further discuss other possible applications of ADMC as well as some outlooks on ADMC.

\section{Background} \label{sec:background}
In this section, we would like to provide some background knowledge for the sake of being self-contained. Specifically, we will introduce some basic knowledge of AD, Monte Carlo gradients estimations, and variational Monte Carlo, which are related to the general theory and applications of ADMC.

\subsection{\bf Automatic differentiation}
Conventional methods of computing gradients of a given function include symbolic and numerical approaches. It is challenging to symbolically compute gradients of complicated functions as deriving the analytical expression of gradient is often nearly impossible. Numerical differentiation approach computes the gradient by finite discretization and thus normally suffers discretization errors. In addition, these two conventional methods encounter more challenges or errors in computing higher order derivatives, especially when the number of input parameters is large.

AD, on the contrary, by tracing the derivatives propagation of primitive operations via chain rules, can render numerically-exact derivatives (including higher order derivatives) for any programs \cite{Bartholomew-Biggs2000a, atlm2018, Margossian2018}. The program is specified by computational graph composed of function primitives. Such directed acylic graph shows the data shape and data flow of the corresponding program.

\begin{figure}[t]\centering
	\includegraphics[width=7cm]{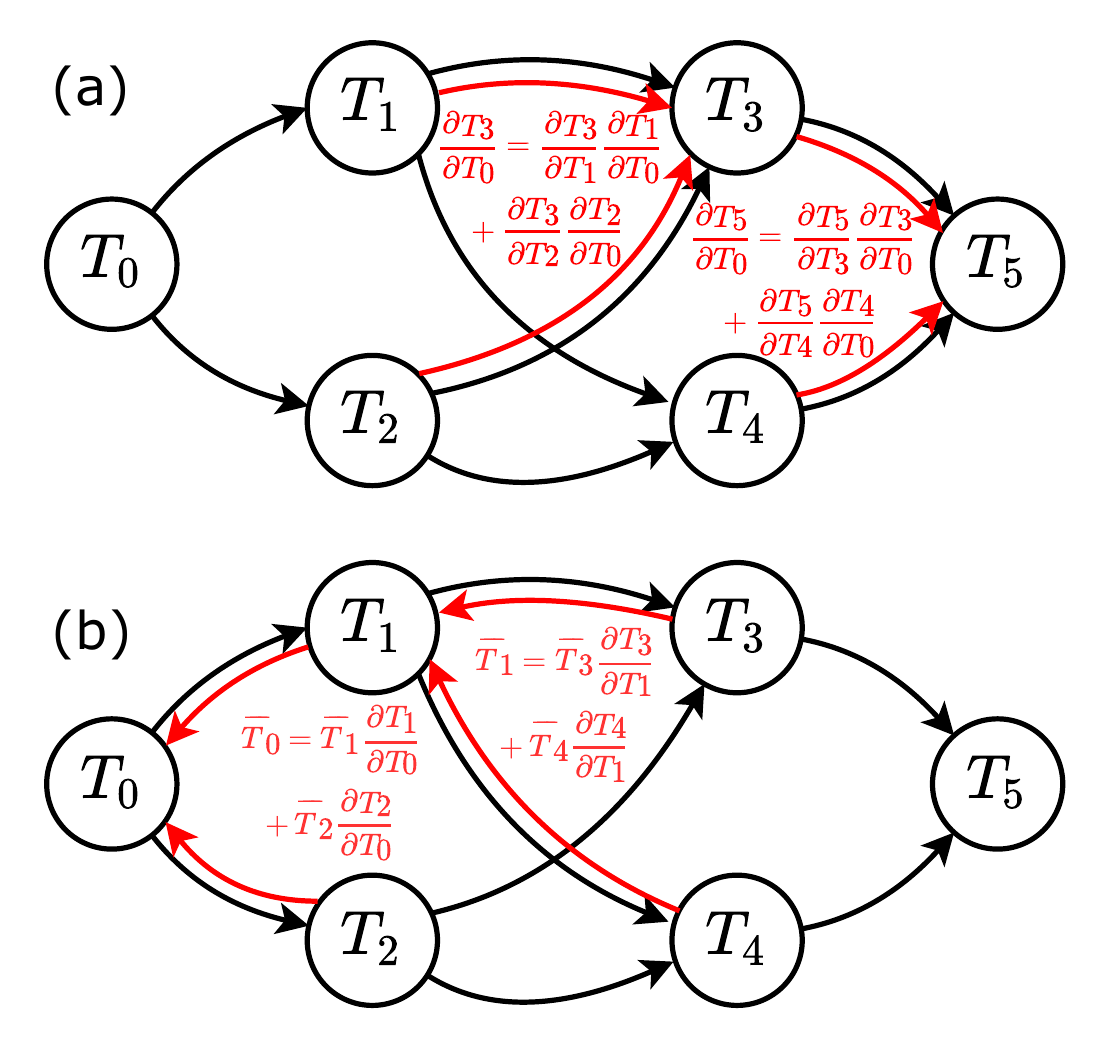}
	\caption{Forward mode (a) and reverse mode (b) automatic differentiation on computational graphs. Black arrows label the forward pass from inputs to outputs. Red arrows represent  forward chain rules in (a) and backpropagation for adjoints in (b). }  \label{admode}
\end{figure}

There are two ways to compute the derivative on the graph with respect to the graph's inputs: the forward AD and backward AD. The forward AD iteratively compute the recursive expression as shown in Fig.~\ref{admode}(a):
\bea
\frac{\p T_i}{\p T_0}=\sum_{T_{i-1}\in \mathrm{parent} \{T_i\}}\frac{\p T_i}{\p T_{i-1}}\frac{\p T_{i-1}}{\p T_{0}},
\label{fad}\eea
where $T_i$ stands for nodes on the computational graph; $T_0$ is the input and $T_n$ the final output. The gradient we aim to obtain is $\frac{\p T_n}{\p T_0}$. Here $\frac{\p T_i}{\p T_{i-1}}$ corresponds to the derivatives of operator primitives $T_i=f(T_{i-1})$, and these derivatives are implemented as AD infrastructures built-in or user customizations.
One drawback of the forward mode AD is that one need to keep track of every derivative $\frac{\p T_i}{\p T_0[i]}$ in the middle of the graph when the input has many parameters, which is normally expensive and inefficient.

Reverse mode AD can avoid the inefficiency encountered by forward mode AD when input parameters are far more than output ones, which is the case of many applications including machine learning and variational Monte Carlo. By defining the adjoint as $\overline{T}_i=\frac{\p T_n}{\p T_i}$. As shown in Fig.~\ref{admode}(b),  reverse mode AD iteratively compute the recursive expression:
\bea
\overline{T}_i=\sum_{T_{i+1}\in \mathrm{child}\{T_{i}\}}\overline{T}_{i+1}\frac{\p T_{i+1}}{\p T_i}.
\label{arg2}\eea
The final aim is to compute $\overline{T}_0$. In this approach of AD, one first computes the output and save all intermediate node values $T_i$ in the forward pass, and then backpropagates the gradients in the reverse pass. This workflow, denoted as backpropagation in machine learning language \cite{Rumelhart1986} is opposite to the forward mode AD where all computations happen in the same forward pass. Reverse mode AD is in particular suitable for scenarios with multiple input parameters and one output value, which is the case of most deep learning setups \cite{Lecun2015} and many MC approaches such as variational Monte Carlo. 

\subsection{\bf Gradients of Monte Carlo expectations}
As explained in the introduction, it is of great importance in various fields to compute gradients of Monte Carlo expectation values: $\nabla_{\vect{\theta}} \langle O(\vect{x},\vect{\theta})\rangle_{p(\vect{x}, \vect{\theta})}$, where $\vect{\theta}$ represents the set of input parameters, $\vect{x}$ label MC configurations, $p(\vect{x},\vect{\theta})$ is the (generally unnormalized) probability distribution, and $\langle O(\vect{x},\vect{\theta})\rangle_{p(\vect{x}, \vect{\theta})}$ is the MC expectation value of $O$ under the probability distribution $p(\vect{x},\vect{\theta})$. We often call $O$ the loss function. Currently, there are two main methods for evaluating MC gradients: score function estimator \cite{Kleijnen1996} (also denoted as REINFORCE \cite{Williams1992}) and pathwise estimator (also denoted as reparametrization trick \cite{Kingma2013}, stochastic backpropagation \cite{Rezende2014}, or ``push out" method \cite{Rubinstein1992}). 
Although the method of pathwise estimator in general gives lower variance for MC gradients estimation, it can only be applied to quite limited settings due to the strict requirements on the differentiability of transformers and probability distributions. Therefore, it is nearly impossible to apply pathwise estimator to evaluate MC gradient sampled from vastly complicated distributions encountered in most physics problems. We thus focus on score function method in the present paper as it is more universal and general-purpose.

The score function estimator is a general-purpose MC gradient estimator with gradient given by 
\bea
\nabla_{\vect{\theta}} \langle O(\vect{x},\vect{\theta})\rangle_p\!=\!\left\langle \nabla_{\vect{\theta}}O(\vect{x},\vect{\theta})+O(\vect{x}, \vect{\theta})\frac{\nabla_{\vect{\theta} }p(\vect{x},\vect{\theta})}{p(\vect{x}, \vect{\theta})}\right \rangle_p,~~~~
\label{score}\eea
where $p$ is a shortcut for $p(\vect{x},\vect{\theta})$. Note that \Eq{score} is quite general 
by taking into account the dependence of the loss function $O$ on the parameters $\vect{\theta}$. To leverage AD on the gradient estimation, 
it is desired to construct an AD-aware version of MC expectation which can correctly obtain the MC gradient itself and its derivatives to all order (including the gradients and all higher derivatives). For normalized probability distribution $p$, the following AD-aware version of MC expectation
\bea
\left\langle \frac{O(\vect{x},\vect{\theta})p(\vect{x}, \vect{\theta})}{\bot(p(\vect{x}, \vect{\theta}))}\right\rangle_p
\label{adscore3}\eea
was proposed \cite{Foerster2018b}. However, for nearly all interesting physics problems, the normalization factor is not known {\it a priori} and probability distribution is usually unnormalized.
 It is of central importance to sample from such unnormalized probability distributions for applications such as computing physical quantities without knowing partition function {\it a priori} or approximating the posterior distributions of latent variables only with knowledge of likelihood and prior. Nevertheless, MC gradient estimation from such unnormalized probability distribution has not been constructed by any previous method. In Sec.~\ref{sec:theory} of the present paper, we develop a general framework and construct the AD-aware objective MC expectation which can correctly obtain both the expectation value and its all higher-order derivatives for unnormalized probability distributions.

\subsection{\bf Variational Monte Carlo in physics}
VMC is a powerful numerical algorithm searching the ground state of a given quantum Hamiltonian based on the variational principle since a physical Hamiltonian has energy bounded from below \cite{Hill1965, Ceperley1977}. By sampling the amplitude of variational wave function $\ket{\psi_{\vect{\theta}}}$, where $\vect{\theta}$ represents the set of variational parameters, one can compute the energy expectation  $E_{\vect{\theta}}=\bra{\psi_{\vect{\theta}}}H\ket{\psi_{\vect{\theta}}}/ \langle\psi_{\vect{\theta}}|\psi_{\vect{\theta}}\rangle$, where the ansatz wave function $\ket{\psi_{\vect{\theta}}}$ is in general not normalized. The energy expectation can be evaluated through MC:
\bea
E_{\vect{\theta}}
=\frac{\sum_\sigma p(\sigma,\vect{\theta})\frac{\langle\sigma \vert H\vert\psi_{\vect{\theta}}\rangle}{\langle \sigma\vert\psi_{\vect{\theta}}\rangle}}{\sum_\sigma  p(\sigma,\vect{\theta})}=\left<E_{\mathrm{loc}}(\sigma,\vect{\theta})\right>_{p(\sigma,\vect{\theta})},
\label{vmc}\eea
where $E_{\mathrm{loc}}(\sigma,\vect{\theta})=\frac{\langle \sigma \vert H\vert \psi_{\vect{\theta}}\rangle}{\langle \sigma\vert\psi_{\vect{\theta}}\rangle}$, $\sigma$ is complete basis of quantum system's Hilbert space, and $p(\sigma,\vect{\theta})= |\langle\sigma\vert\psi_{\vect{\theta}}\rangle|^2$ is the probability distribution. 
Note that the probability distribution $p(\sigma,\vect{\theta})$ is in general unnormalized since the ansatz wave function $\psi_{\vect{\theta}}(\sigma)=\langle\sigma|\psi_{\vect{\theta}}\rangle$ is in general unnormalized (as it is usually challenging to normalize the ansatz wave function due to complicated wave function structure). Since $E_{\vect{\theta}}$ depends on variational parameters $\vect{\theta}$, one thus can in principle optimize $E_{\vect{\theta}}$ obtained by \Eq{vmc} against $\vect{\theta}$, giving rise to the optimal ground state energy and wave function within the ansatz.

Stochastic gradient descent (SGD) is {\textit{de facto}} for optimizations in machine learning \cite{Robbins1951, Kiefer1952, Bottou2018a} and can also be employed in computational physics such as optimization in VMC \cite{Harju1997c}. There are various generalizations beyond vanilla SGD optimizers by considering momentum and adaptive behaviors, amongst which Adam \cite{Kingma2014} is one common optimizer in training neural networks. Natural gradient descent, a concept emerged from information geometry, is one of the optimization techniques where the local curvature in distribution space defined by neural networks has been considered \cite{Amari1998, Pascanu2013a, Martens2014, Karakida2019a}. Efficient approximations on natural gradient have also been investigated such as FANG \cite{Grosse2015} and K-FAC \cite{Martens2015, Grosse2016, Ba2017, Martens2018, Osawa}. For optimization problem such as VMC, gradient descent and natural gradient descent methods can be applied where various machine learning techniques can be utilized to boost VMC. Natural gradient optimization is exactly equivalent to stochastic reconfiguration (SR) method \cite{Sorella1998a, Sorella2001} in VMC \cite{Nomura2017, Glasser2018b, Pfau2019a,Park2019}. 

Recently, there were various studies focusing on using restricted Boltzmann machine (RBM) or related neural networks as the ansatz wave function for quantum systems composed of spins \cite{Carleo2017,Deng2017a, Carleo2018, Cai2018a, Kochkov2018,Kaubruegger2018, Glasser2018b,Choo2018, Liang2018, Vieijra2019, Yang2019, He2019}, bosons \cite{Nomura2017, Saito2017a, McBrian2019}, and fermions \cite{Pfau2019a,Hermann2019a}. In previous studies, to incorporate such wave function ansatz into the framework of VMC, one either computes all derivatives ${\nabla_{\vect{\theta}} \psi_{\vect{\theta}}(\sigma)}$ analytically when the neural network ansatz is simple enough \cite{Carleo2017} or applies AD on the wave function to compute ${\nabla_{\vect{\theta}} \psi_{\vect{\theta}}(\sigma)}$ \cite{Yang2019}, and then estimate the gradient ${\nabla_{\vect{\theta}} E_{\vect{\theta}} }$ by MC sampling given by
\eq{\nabla_{\vect{\theta}} E_{\vect{\theta}}=2 \Re[\langle\frac{\nabla{\psi_\sigma}^*}{\psi_\sigma^*}E_{\mathrm{loc}}\rangle- \langle E_{\mathrm{loc}}\rangle\langle\frac{\nabla{\psi_\sigma}^*}{\psi_\sigma^*}\rangle],}{vmcd}
where $\langle\cdot \rangle$ denotes MC sampling of configurations $\sigma$ from probability distribution $\abs{\psi(\sigma)}^2$. 
However, applying AD directly on the energy expectation $E_{\vect{\theta}}=\langle\psi_{\vect{\theta}}|H|\psi_{\vect{\theta}}\rangle$ to obtain the gradient ${\nabla_{\vect{\theta}} E_{\vect{\theta}}}$, the most intuitive way to optimize the ground state, is still lacking partly due to the lack of AD technique for MC expectations sampled from unnormalized probability distributions. With the introduction of ADMC in this work, we can implement AD-aware VMC, which is much more straightforward and easy to implement by directly optimizing the energy expectation without any analytical derivation on derivatives for MC expectations or wave functions, which we call ``end-to-end'' ADVMC.

\section{Theory} \label{sec:theory}
In this section, we will present the general theory of the ADMC which enables infinite-order automatic differentiation on MC expectations with unnormalized probability distributions. We shall show the detailed derivations on the general theory.

\subsection{\bf Detach function}
We first introduce detach function $\bot(x)$ which features $\bot(x)=x$ and $\frac{\p \bot(x)}{\p x}=0$. Here we list some basic formula in terms of detach functions utilized later: $f(\bot(x))=\bot(f(x))$, $\bot(\bot(x)) = \bot(x)$, $\bot(x+y)=\bot(x)+\bot(y)$, and $\bot(xy) =\bot(x)\bot(y)$. The detach function can be easily implemented and simulated in modern machine learning frameworks (it corresponds to \inlinecode{python}{stop\_gradient} in TensorFlow \footnote {See \url{https://github.com/tensorflow/tensorflow}} and \inlinecode{python}{detach} in PyTorch {\footnote {See \url{https://github.com/pytorch/pytorch}}}).
We call this function primitive as detach function in this work. This weird looking function has natural explanation in the context of machine learning, especially in terms of computational graph. Such operator corresponds to node in the graph which only pass forward values while stop the back propagation of gradients.

By utilizing detach function, we can construct functions whose derivatives of each order is not related. For example, the function $O(x)=x-\bot(x)$ equals to $0$ irrespective of $x$ although its first-order derivative is $1$. The detach function is mathematically sound as we shall prove a completeness theorem for the detach function below.

\begin{theorem}
For any ``weird" function, whose value and every order of derivatives are defined separately, it can always be expressed by ``normal" functions with detach function $\bot$.
\end{theorem}
{\underline {Proof.}}
For a function $\mathcal{F}(x)$ whose each order of derivatives are defined as $\mathcal{F}^{(n)}(x)=h_n(x)$, the construction with $\bot$ is:
\bea
\mathcal{F}(x)=\sum_{n=0}^{\infty} \frac{1}{n!} h_n(\bot(x))(x-\bot(x))^n.
\label{comthm}\eea
When translated into TensorFlow language, Theorem 1 states that every function defined with \inlinecode{python}{tf.custom_gradient} can be instead defined with \inlinecode{python}{tf.stop\_gradient}.

\begin{corollary}
For a function with multiple variate input $\mathcal{F}(x_1,...x_m)$ whose derivatives $\mathcal{F}^{(n_1...n_m)}$ are defined separately, irrelevant from the original function, it can always be expressed by ``normal" functions together with single variate detach function $\bot$.
\end{corollary}
The corollary above is obvious by considering similar Taylor expansion construction as \Eq{comthm}.

The introduction of imaginary number $i$ enlarges the meaning of the equal sign by twice the equivalence relation: $x=y\Leftrightarrow \Re(x)=\Re(y)$ and $\Im(x)=\Im(y)$. Similarly, with the introduction of detach function, the equal sign are enlarged as infinite independent equivalence relations: $f(x)=g(x)\Leftrightarrow  \bot(f^{(n)}(x))=\bot(g^{(n)}(x)), (n=0,1,2,\cdots)$. The conventional ``equal" is reexpressed as one relation ($n=0$) of the above series:  $\bot(f(x)) = \bot(g(x))$.

\subsection{\bf ADMC}

We are ready to construct a general theory for MC expectation which can render AD to correctly obtain its directives at all order (including the zeroth-order derivative, the expectation itself). We employ the extended score function method to enable AD on MC expectations for any complicated distribution, both normalized and unnormalized. Theorem 2 below is the central theoretical result of the present paper.

\begin{theorem}
The following MC estimator of $\langle O(\vect{x},\vect{\theta})\rangle_{p}$
\bea
\left\langle O(\vect{x}, \vect{\theta})\right\rangle_{p}
	=\frac{ \left   \langle \frac{p}{\bot(p)}O(\vect{x}, \vect{\theta})\right\rangle_{\bot(p)}}
{\left\langle\frac{p}{\bot(p)}\right\rangle_{\bot(p)}}
\label{mcadobj}\eea
is automatic differentiable to all order and works for both normalized and unnormalized probability distribution $p=p(\vect{x},\vect{\theta})$.
\end{theorem}

To prove Theorem 2, we first introduce the following lemma:
\begin{lemma}
For both normalized and unnormalized probability distribution $p=p(\vect{x},\vect{\theta})$,
\eq{\sum_{\vect{x}\in S(p)} \frac{p}{\bot(p)} \doteq \frac{Z}{\bot (Z)}.}{lemma}	
\end{lemma}
Here $Z$ is the shortcut for partition function $Z_{\vect{\theta}}=\sum_{\vect{x}\in \mathrm{all}}p(\vect{x}, \vect{\theta})$ with $\vect{x}\in \mathrm{all}$ representing the summation over all configurations $\vect{x}$. $\sum_{\vect{x}\in S(p)}$ denotes the average obtained through MC sampling according to the probability distribution $p$. Note that, for brevity, we omit the $\frac{1}{N_s}$ factor before the MC sum $\sum_{\vect{x}\in S(p)}$ in \Eq{lemma} and hereafter; the sum should be understood as the average $\frac{1}{N_s}\sum_{\vect{x}\in S(p)}$ where $N_s$ is the number of sample configurations. In \Eq{lemma} and hereafter, ``$\doteq$'' means that it is the same as the equal sign since MC estimation can be made exact in the limit of large $N_s$. The equal sign also makes sense in any order derivatives. Therefore, to prove the lemma we just need to demonstrate the following formula:
\bea
\bot\left(\sum_{\vect{x}\in S(p)} \nabla_{\vect{\theta}}^{(n)} \frac{p}{\bot(p)}  \right) \doteq \bot\left(\nabla_{\vect{\theta}}^{(n)} \frac{Z}{\bot(Z)}\right),
\label{complex1}\eea
where $\nabla^{(n)}_{\vect{\theta}}$ is a shortcut for $\nabla^{(n_1,\cdots,n_m)}_{\theta_1,\cdots,\theta_m}$, $n_j=0,1,2,\cdots$.

For $n=0$, the equation is simply true since both sides give $1$. For arbitrary $n$, it is straightforward to show that
\bea
\bot\left(\sum_{\vect{x}\in S(p)}\frac{\bot(\nabla^{(n)} p)}{\bot(p)}\right)&\doteq& \sum_{\vect{x}\in \mathrm{all}} \frac{\bot(p)}{\bot(Z)} \frac{\bot\nabla^{(n)} p}{\bot(p)} \nonumber\\
 &=&\frac{\bot\nabla^{(n)} \sum_{\vect{x}\in \mathrm{all}} p}{\bot Z} \nonumber \\
    &=&\frac{\bot(\nabla^{(n)} Z)}{\bot(Z)},
\label{complex2}\eea
which finishes the proof of the lemma.

The proof of the lemma above can be significantly simplified. With the enlarged meaning of equal sign, each order of derivatives automatically equal as long as expressions with detach function are accordingly considered. In other words, to prove that some relation holds true for any order derivatives $\bot(f^{(n)}(x))=\bot(g^{(n)}(x))$, $(n=0,1,2,\cdots)$, we only need to prove that $f(x) = g(x)$ is true. This simplification is the power of detach function. The proof of the lemma can be simplified as:
\bea
\sum_{\vect{x}\in S(p)}\frac{p}{\bot{(p)}} \doteq \sum_{\vect{x}\in \mathrm{all}}\frac{\bot{(p)}}{\bot{(Z)}}\frac{p}{\bot{(p)}} =
	\frac{\sum_{\vect{x}\in \mathrm{all}} p}{\bot(Z)}=\frac{Z}{\bot{(Z)}}.~~~~~
\label{simple}\eea
Note that the simplification from the involved proof in \Eq{complex1} and \Eq{complex2} to the neat one in \Eq{simple} reflects the brevity and power of detach function and its algebra.

Now we are ready to prove Theorem 2. Proving this theorem is equivalent to show the following equation:
\bea
\bot\left(  \nabla^{(n)}_{\vect{\theta}}
	\frac{\langle  \frac{p}{\bot(p)} O \rangle_{\bot(p)}}
	{\langle\frac{p}{\bot(p)}\rangle_{\bot(p)}}  \right)\doteq \bot\left(\nabla^{(n)}_{\vect{\theta}} \sum_{\vect{x}\in \mathrm{all}}p\frac{O}{Z}  \right),
\label{arg2}\eea
where $O$ is the shortcut for $O(\vect{x}, \vect{\theta})$. 
Note that, in the average $\langle\cdot \rangle_{\bot (p)}$, the probability distribution $\bot(p)$ is the background and is not involved in derivatives. Based on the spirit of detach function algebra, it is enough to show:
\bea
\sum_{\vect{x}\in S(p)} \frac{p}{\bot(p)} O \big/\sum_{\vect{x}\in S(p)} \frac{p}{\bot (p)} \doteq \sum_{\vect{x}\in \mathrm{all}} p\frac{O}{Z}.
\label{proof}\eea
By utilizing the lemma in \Eq{lemma} and observing the fact that $Z_{\vect{\theta}}$ is independent of $\vect{x}$, it is straightforward to prove \Eq{proof} as follows:
\bea
\sum_{\vect{x}\in S(p)} \frac{p}{\bot(p)} O \big/\sum_{\vect{x}\in S(p)} \frac{p}{\bot (p)} &\doteq& \sum_{\vect{x}\in \mathrm{all}} \frac{\bot(p)}{\bot(Z)}\frac{pO}{\bot(p)}\big/(\frac{Z}{\bot(Z)}) \nonumber \\
&=&\sum_{\vect{x}\in \mathrm{all}}p\frac{O}{Z}. \nonumber 
\eea
This finishes the proof of Theorem 2, which is the central result of the present paper. We believe Theorem 2 can provide endless opportunities to build applications combining AD infrastructure with MC algorithms.

We emphasize that Theorem 2 is general and applies for both normalized and unnormalized probability distribution $p$. For the case of normalized distribution $\sum_{\vect{x}\in \mathrm{all}} p =1$, we obtain $\sum_{\vect{x}\in S(p)}\frac{p}{\bot(p)}\doteq \sum_{\vect{x}\in \mathrm{all}}\bot(p) \frac{p}{\bot(p)}=\sum_{\vect{x}\in \mathrm{all}}p=1$.
Then, \Eq{mcadobj} in Theorem 2 can be simplified to $\left\langle O\right\rangle_{p}=\left\langle \frac{p}{\bot(p)}O\right\rangle_{\bot(p)}$, which is the MC estimator applicable only for the case of normalized probability distribution. For nearly all interesting applications with unnormalized probability distribution $p$, Theorem 2 is the correct one to use, as we demonstrate in the applications below.

It is worth to provide heuristic explanation for Theorem 2. Through discretizing the parameters $\vect{\theta}$ in numerical differentiations, rigorous MC gradient can be obtained in the limit of zero distretizing intervals. Specifically, to get gradients at $\vect{\theta_0}$, one can directly compute MC expectations of $O$ by sampling separately from $p(\vect{\theta})$ and from $p(\vect{\theta}_0)$, with $\vect{\theta}$ very close to $\vect{\theta_0}$. However, it is highly inefficient to sample separately from $p(\vect{\theta})$ and from $p(\vect{\theta}_0)$ distributions. As $\vect{\theta}$ is close to $\vect{\theta_0}$ (in the limit $\vect{\theta}\rightarrow\vect{\theta_0}$), one can actually reuse the samples from $p(\vect{\theta}_0)$ to evaluate the expectation at $\vect{\theta_0}$:  
\bea
&&\langle O(\vect{x},\vect{\theta})\rangle_{p(\vect{\theta})}=\sum_{\vect{x}\in \mathrm{all}} p(\vect{\theta})O(\vect{x}, \vect{\theta})\big/\sum_{\vect{x}\in \mathrm{all}} p(\vect{\theta})\nonumber\\
&&=\frac{1}{N}\sum_{\vect{x}\in \mathrm{all}} p(\vect{x},\vect{\theta}_0)\frac{p(\vect{x},\vect{\theta})}{p(\vect{x},\vect{\theta}_0)}O(\vect{x}, \vect{\theta})\big/\sum_{x\in \mathrm{all}} p(\vect{x},\vect{\theta}_0)\frac{p(\vect{x},\vect{\theta})}{p(\vect{x},\vect{\theta}_0)}\nonumber\\ 
&&=\left<\frac{p(\vect{x},\vect{\theta})}{p(\vect{x},\vect{\theta}_0)}O(\vect{x}, \vect{\theta})\right>_{p(\vect{\theta}_0)}\big/\left< \frac{p(\vect{x},\vect{\theta})}{p(\vect{x},\vect{\theta}_0)} \right>_{p(\vect{\theta}_0)}.
\label{naive}
\eea
By comparing \Eq{naive} with \Eq{mcadobj}, one can observe the parallel relations between them and understand the physical rational behind detach functions: when evaluating derivatives only $\vect{\theta}$ changes while $\vect{\theta}_0$ fixed, all terms related to $\vect{\theta}_0$ are wrapped with detach function $\bot$ in the exact form \Eq{mcadobj} in Theorem 2. 

Finally, we make a note on implementation. For numerical stability, $\ln p$ instead of $p$ is in general referenced and the AD version of MC estimator for generic probability distribution $p$ is then given by:
\bea
\langle O\rangle_p = \frac{\langle\mathrm{exp}(\ln p-\bot(\ln p)) O\rangle_{\bot p}}{\langle\mathrm{exp}(\ln p -\bot(\ln p))\rangle_{\bot p}}.
\label{arg2}\eea
From computational graph implementation perspective, $p$ is never explicitly calculated since the numerical value of $\mathrm{exp}(\ln p -\bot(\ln p))$ is exactly one. Therefore, ADMC approach using $\ln p$ is automatically free from the numerical instability encountered in approaches directly using $p$.

\subsection{\bf Fisher information matrix and KL divergence in ADMC}

For the optimization method of natural gradient descent, the parameters $\vect{\theta}$ are updated in the following way: $\Delta \vect{\theta} =-\lambda F^{-1} \nabla_{\vect{\theta}} O_{\vect{\theta}}$, 
where $F$ is the Fisher information matrix (FIM), $\lambda$ is the learning rate and $O_{\vect{\theta}}=\langle O(\vect{x},\vect{\theta})\rangle_p$. FIM is of great importance in numerical optimization and is defined as
\bea
F_{ij}&=&\left< \nabla_i \ln \frac{p}{Z} \nabla_j \ln \frac{p}{Z}\right>_p, 
\eea
where $i,j$ represent $\theta_i,\theta_j$.  FIM is also the Hessian (with respect to $\vect{\theta'}$) of KL divergence between $p(\vect{x},\vect{\theta})$ and $p(\vect{x}, \vect{\theta}')$ with $\vect{\theta}'$ approaching $\vect{\theta}$. Hence, it defines the local curvature in distribution space.

In the following, we derive useful formulas related to FIM with unnormalized probability distribution $p$ in the context of ADMC.
For unnormalized $p$, the expectation of score function is not zero and it is given by:
\bea
\langle \nabla_{\vect{\theta}}\ln p\rangle_p=\frac{1}{Z}\sum_{\vect{x}\in \mathrm{all}} p\frac{\nabla_{\vect{\theta}} p}{p}=\frac{\nabla_{\vect{\theta}} Z}{Z}=\nabla_{\vect{\theta}} \ln Z.
\label{arg2}\eea
Then, the FIM for unnormalized $p$ can be defined as
\bea
F_{ij}
=\langle \nabla_i \ln p\nabla_j \ln p\rangle_p-\langle \nabla_i \ln p\rangle_p\langle\nabla_j \ln p\rangle_p.
\label{fimu}\eea

 To apply AD approach, we can obtain FIM through the KL divergence whose Hessian is FIM. The AD-aware KL divergence is given by 
\bea
\mathrm{KL}\left(\bot(\frac{p}{Z})\vert \frac{p}{Z}\right)&=&\ln \frac{Z}{\bot(Z)}-\left \langle \ln \frac{p}{\bot{(p)}}\right \rangle_{\bot(p)}\nonumber\\
&=&\ln\left \langle\frac{p}{\bot(p)}\right\rangle_{\bot(p)}-\left\langle \ln \frac{p}{\bot{(p)}} \right\rangle_{\bot(p)},~~~~~~
\label{adfimu}\eea
where the second equation is due to \Eq{lemma} in Lemma 1. Therefore, for any unnormalized $p$, we can construct object function as \Eq{adfimu} and compute Hessian of it by ADMC. This approach is preferable than direct estimation from \Eq{fimu} in some scenarios. (see the SM \footnote{See Supplemental Materials for details. The SM of this work includes: 1. Detailed discussion on the advantages and implementations for obtaining FIM within ADMC framework. 2. Detailed derivation and implementation details of end-to-end VMC for positive valued and general complex valued wave function case.} for details)

Following the path of \Eq{adfimu}, we could further derive the AD-aware formula for general KL divergence with unnormalized probability $p,q$ parameterized by $\vect{\theta}$ as

\eq{\mathrm{KL}(p_{\vect{\theta}}\vert q_{\vect{\theta}})=\ln \frac{\left\langle\frac{p}{\bot{p}}\frac{q}{p} \right\rangle_{\bot p}}{\left\langle\frac{p}{\bot{p}}\right\rangle_{\bot p}}-\frac{\left\langle \frac{p}{\bot{p}}\ln \frac{q}{p}\right\rangle_{\bot{p}}}{\left\langle \frac{p}{\bot p}\right\rangle_{\bot{p}}}.}{arg2}

\subsection{End-to-end ADVMC}
As discussed in Sec. \ref{sec:background}, VMC is an important approach attempting to find the ground state wave function of a Hamiltonian by optimizing parametrized wave functions. Here we describe how to implement end-to-end VMC with ADMC, which we call ADVMC. We shall focus on the case where ansatz wave functions are positively valued. For the general case of complex-valued ansatz wave functions, ADVMC can also be implemented. (see the SM \cite{Note3} for details)

As in \Eq{vmc}, the energy expectation value $E_{\vect{\theta}}=\bra{\psi_{\vect{\theta}}}H\ket{\psi_{\vect{\theta}}}$ of Hamiltonian $H$ associated with the wave function $\psi_{\vect{\theta}}(\sigma)=\langle\sigma\ket{\psi_{\vect{\theta}}}$ can be evaluated through Monte Carlo sampling \bea
E_{\vect{\theta}}=\left<E_{\mathrm{loc}}(\sigma,\vect{\theta})\right>_{p(\sigma,\vect{\theta})}, 
\eea
where $p(\sigma,\vect{\theta})=|\psi_{\vect\theta}(\sigma)|^2$ is usually unnormalized probability distribution. To optimize (minimize) $E_{\vect{\theta}}$ using gradient-based approach, we need to evaluate the gradients with respect to variational parameters $\vect \theta$:
\bea
\nabla_{\vect \theta}\left<E_{\mathrm{loc}}(\sigma,\vect{\theta})\right>_{p(\sigma,\vect{\theta})}.
\label{VMC_expect}\eea
It is clear that $E_{\mathrm{loc}}(\sigma,\vect{\theta})$ in VMC plays a similar role as $O(\vect{x},\vect{\theta})$ in MC discussed earlier. It is natural to integrate AD into VMC so that an end-to-end ADVMC can be constructed.
The ADVMC version of the energy estimator can be constructed as follows: 
\bea
\left \langle E_{\mathrm{loc}}(\sigma,\vect{\theta})\right \rangle_p = \Re\frac{\left\langle\frac{p}{\bot p}E_{\mathrm{loc}}(\sigma,\vect{\theta})\right\rangle_{\bot{p}}}{\left\langle\frac{p}{\bot p}\right\rangle_{\bot p}},
\label{eloc}\eea
where $\Re$ guarantees that the energy estimator is real. 

Taking account of the variance reduction trick, the ADVMC energy estimator for real wave function can also be constructed as \cite{Note3}:
\bea
	\left \langle E_{\mathrm{loc}}(\sigma,\vect{\theta})\right \rangle_p =\Re\frac{\left<\frac{\psi^2}{\bot(\psi^2)}\bot (E_{\mathrm{loc}}(\sigma,\vect{\theta}))\right>_{\bot(p)}}
		{\left<\frac{\psi^2}{\bot(\psi^2)}\right>_{\bot{(p)}}}.
\label{VMC_estimator}\eea
Actually, the objective in \Eq{VMC_estimator} has better performance compared with the original estimator in \Eq{eloc} since $E(\sigma,\vect{\theta})$ is detached in \Eq{VMC_estimator} and no further backpropagations behind this node are needed. Note that \Eq{VMC_estimator} as the estimator of $E_{\vect{\theta}}$ can only reproduce first-order derivative in the framework of AD, while the original estimator in \Eq{eloc} is correct for all order derivatives. (see the SM \cite{Note3} for details)

The end-to-end ADVMC framework is universal and easy to implement. Instead of computing derivatives of wave functions and plugging the results into the formula of energy gradients by hands as conventional VMC approaches do in \Eq{vmcd}, the end-to-end ADVMC optimizes the energy expectation directly and leaves all remaining work to machine learning infrastructure. Analytic and implementation works can be done automatically with AD infrastructure, vectorization/broadcast mechanism, builtin optimizers and GPU acceleration provided by standard ML framework. For different quantum models, the only difference is different $E_{\mathrm{loc}}(\sigma,\vect{\theta})$. After implementing $E_{\mathrm{loc}}$, we can bring it into \Eq{VMC_estimator} as AD-aware energy estimator.  Then, we can use AD to compute the gradients and gradient-based optimizer to optimize the energy. 

Besides SGD-based optimizers, natural gradient optimizers (SR methods) can also be incorporated into AD framework.
In the context of VMC, the optimization method of natural gradient descent updates the variational parameters as follows: $\Delta \vect{\theta} = -\lambda F^{-1} \nabla_{\vect{\theta}} E_{\vect{\theta}}$
where $F$ can be obtained by Monte Carlo:
\bea
F_{ij}= \Re\left[\left<\frac{\p_i \psi^\ast}{\psi^\ast}\frac{\p_j\psi}{\psi}\right>_p
-\left<\frac{\p_i \psi^\ast}{\psi^\ast}\right>_p \left<\frac{\p_j \psi}{\psi}\right>_p\right],~~
\label{srf}\eea
where $\psi$ is a shortcut for $\psi_{\vect{\theta}}(\sigma)$ and dependence on parameters $\vect{\theta}$ is implicit. Note that \Eq{srf} is connected to \Eq{fimu} when the distribution $p=|\psi|^2$ and $\psi$ is real. The relation between FIM and SR method with complex wave functions can also be analyzed by generalizing KL divergence in complex distribution case.  (see the SM \cite{Note3} for details)

\section{Applications} \label{sec:application}
The general theory of ADMC we presented above has broad applications, including achieving high accuracy and efficiency in studying interesting many-body interacting models in physics. As we mentioned earlier, by introducing ADMC, we can leverage not only AD but also other powerful features of machine learning frameworks to traditional Monte Carlo. AD together with vectorization, GPU acceleration, and state-of-the-art optimizers can build faster and more capable Monte Carlo applications to study challenging issues in statistics and physics. Here we present two explicit ADMC's applications in studying interacting many-body systems \footnote{Code implementation of the applications can be found at \url{https://github.com/refraction-ray/admc}} where comparable or higher accuracy can be achieved comparing with previous studies using RBM-based VMC, and tensor network methods.

\subsection{\bf Fast search of phase transitions by ADMC}
For many-body systems in physics, it is among central interest to find distinct phases and phase transitions between them. We shall show that ADMC can provide a general and efficient way to find phase transitions in many-body interacting models. At a given phase transition, certain quantities such as specific heat and ordering susceptibility reach a maximal value. This naturally enables ADMC to locate the phase transition in a fast and efficient way by searching for the maximum. A phase transition can occur when certain parameter such as temperature, pressure, and magnetic field is tuned across a critical value. ADMC can efficiently find the critical value of tuning parameter, such as transition temperature.   

For concreteness, we shall use ADMC to find the transition temperature of the 2D Ising model on square lattice as an example, although the approach we shall present is general and can be applied to both classical or quantum models. For quantum models, we call the corresponding AD approach as ``AD quantum Monte Carlo" (ADQMC). The 2D Ising model is given by $H=-\sum_{\langle ij\rangle}J \sigma_i\sigma_j$,
where $\sigma_i=\pm 1$ is the Ising spin on site $i$ of the square lattice and we take $J=1$ as the energy unit. It is well-known that there is a phase transition at critical temperature $T_c$ below which the system orders spontaneously \cite{Onsager1944}.
The Ising model can be MC sampled with unnormalized probability distribution $p(\sigma,T)=\mathrm{exp}(-H(\sigma)/T)$. 
As specific heat reaches a maximal value at the phase transition, conventional MC methods usually compute specific heat for many temperature points and then locate the peak of specific heat curve as phase transition. In these conventional approaches, 
it requires analytical derivation of the formula for specific heat since MC sampling usually cannot compute the specific heat directly. It is relatively simple for specific heat due to the fluctuation-dissipation theorem, {\it i.e.} $C_v(T)=(\langle H^2\rangle_p-\langle H\rangle^2_p)/T$. However, it is generally challenging to analytically derive quantities such as gradient or higher order derivatives of physical quantities. 

\begin{figure}[t]\centering
	\includegraphics[width=7cm]{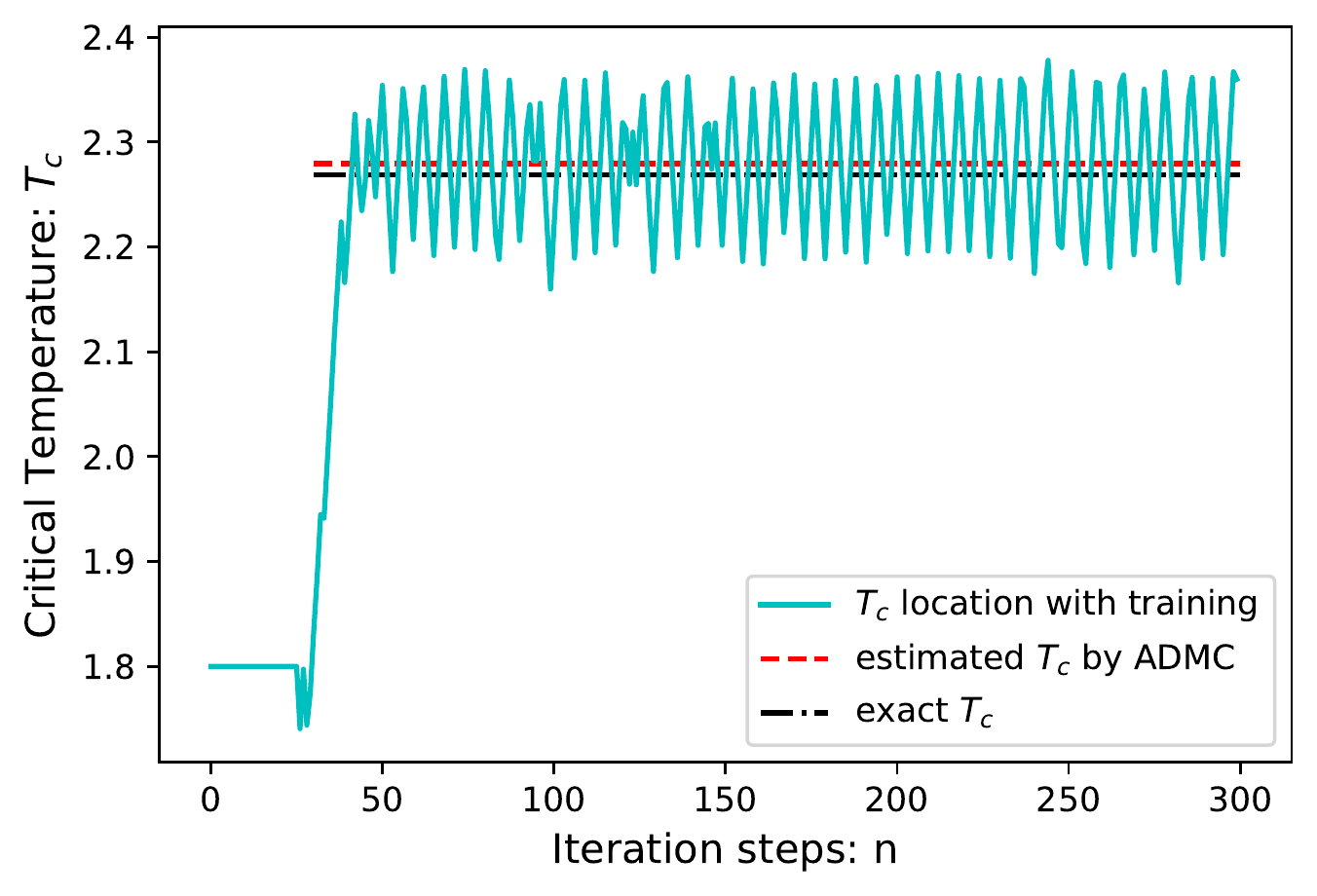}
	\caption{Fast search for critical temperature $T_c$ by the ADMC approach for Ising model of lattice size $50\times 50$ . The obtained expectation value of $T_c$ from training is $2.279$, about 0.4\% off from the exact value $2.269$ \cite{Onsager1944}. Considering the short training time and finite size effect, this is a very good estimation on $T_c$ and more accurate results can be obtained by larger systems size and smaller learning rate. }  \label{isingtc}
\end{figure}

ADMC provides a general way to search for phase transition by directly using the specific heat $C_v(T)$ or other physical quantities as the objective function, which avoids the drawback of conventional MC approaches mentioned above. With the help of ADMC, we can find the peak of the specific heat curve much faster and more efficient. Without the knowledge on the fluctuation-dissipation theorem, we can find the location of the peak very accurately with the total computation time which is orders of magnitude faster. In ADMC, we first directly compute energy using the AD-aware version of MC energy estimator as \Eq{mcadobj} and then, following the spirit of SGD, we update temperature (starting from any $T_0$) based on the second-order derivative of MC expectation energy in every few MC updates:
\bea
\Delta T \propto \frac{\p C_v}{\p T}=\frac{\p^2 \langle \frac{p}{\bot(p)}H\rangle/\langle \frac{p}{\bot{p}}\rangle}{\p T^2}.
\label{arg2}\eea

Although the number of MC updates in each round of temperature update is small rendering noisy estimation of specific heat, such noisy gradient estimator can still converge to $T_c$ very quickly. This is the essence of SGD: noisy gradient estimation might lead to better and faster convergence to the minimum or maximum. This is also why mini-batch gradient estimate is used in general neural network training; for instance, one MC sample each pass in training of variational auto-encoder \cite{Kingma2013} and CD-1 algorithm in RBM training \cite{Hinton2012} work quite well. Following the same philosophy, we can combine SGD into ADMC framework applied here. Specifically, to maximize some MC expectation values against variational parameters $\vect{\theta}=\mathrm{argmax}_{\vect{\theta}} \langle O(\vect{x}, \vect{\theta})\rangle_{p(\vect{x}, \vect{\theta})}$, we may obtain noisy estimation on the gradients by doing few MC update steps. Such noisy estimation on the gradients can render stable and faster optimizations if learning rate is small enough. 

Moreover, one can also utilize third order derivative of expectation energy and apply Newton method to update the temperature, which convincingly shows the value of {\it infinitely} automatic differentiable MC estimators.

 In terms of implementation, we also combine vectorization into the ADMC workflow above, which takes Markov chain as one of the extra dimension for spin configuration tensors, enabling MC simulation on tens of thousands Markov chains simultaneously. Such vectorization scheme is highly efficient compared to conventional parallel schemes, such as one Markov chain per CPU core. Besides, GPU supports such vectorization very well, providing further speed up. The combination of SGD and vectorized Wolff algorithm leads relatively accurate estimation on $T_c$ in just a few seconds. 

We emphasize that the approach we present here is general and can be straightforwardly generalized to other classical or quantum models, where fast estimation on critical values is desired. The knowledge of critical values is helpful to reduce unnecessary calculations on data points deeply in phases and renders fast search of phase transitions in interacting many-body systems. 

\begin{figure}[t]
	\includegraphics[width=0.45\textwidth]{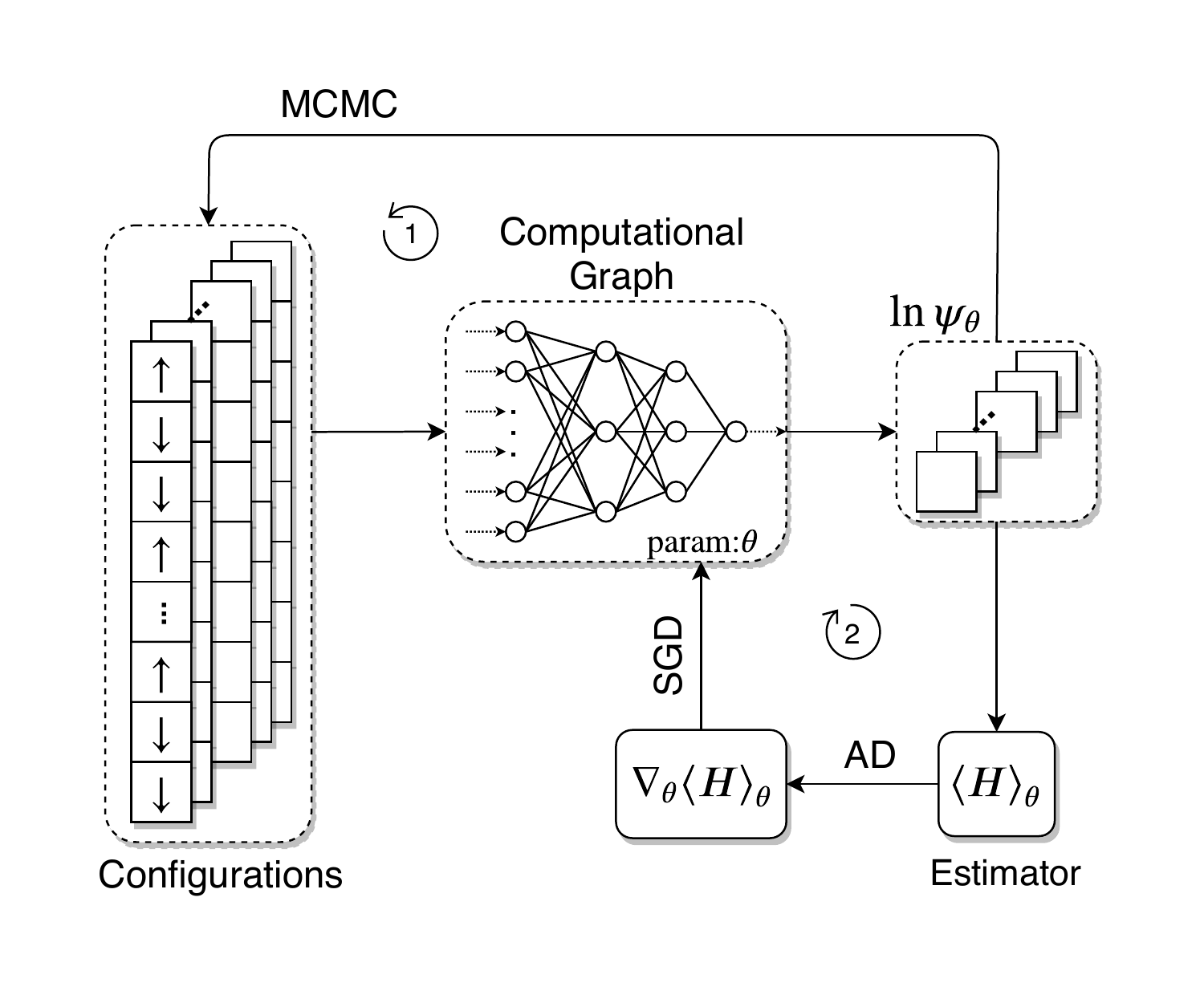}
	\caption{Schematic illustration of the end-to-end ADVMC.
		 Configurations (spins) are vectorized in an extra dimension as different Markov chains. A computational graph is constructed to give the logarithm of wave function $\ln \psi_{\vect \theta}$ which is also vectorized. Loop 1 is the conventional MCMC approach for updating the configurations according to Metropolis-Hasting algorithm. Loop 2 is ADMC approach to evaluate the AD-aware energy estimator and update the parameters $\vect{\theta}$ by optimizers. One iteration of our algorithm include many (often size of the system) configuration updates (loop1)  to decrease the autocorrelation and one step of parameter update (loop2).}  \label{Schematic}
\end{figure}

\subsection{\bf Accurate search of ground states by ADVMC}
The integration of AD with VMC provides a powerful tool to accurately study ground states of many-body quantum models in one and higher dimensions. Especially, ADVMC can study generic quantum models (including those models with frustration) in two and higher dimensions using general neural-network states as ansatz wave functions.

The workflow of the end-to-end ADVMC is sketched in \Fig{Schematic}. In ADVMC algorithm, we can take advantage of the vectorization technique to watch and update thousands of independent Markov chains in the parallel fashion. As shown in \Fig{Schematic}, the (spin) configurations of different Markov chains are vectorized in a new dimension of size $n_{mc}$ (the number of Markov chains). The configurations are sent to an arbitrary computational graph with variational parameters $\vect\theta$ where the logarithm of wave function $\ln \vect\psi_{\vect\theta}(\sigma)$ is evaluated as the output. Computation graph can be constructed by mean-field wave functions with Jastrow factors, matrix product state \cite{White1992, Schollwock2011}, deep neural networks or any other programs with variational parameters and one scalar output. $\ln \vect\psi_{\vect\theta}$ also has an extra dimension with the same size as $n_{mc}$. In evaluating the computational graph, the extra dimension behaves as batch dimension in ML language which can be easily taken care of using broadcast technique supported by ML. With the knowledge of wave function amplitudes, we can update the configurations using MCMC method to make them satisfying the distribution given by computational graph wave function ansatz.

\begin{figure}[t]
	\includegraphics[width=0.35\textwidth]{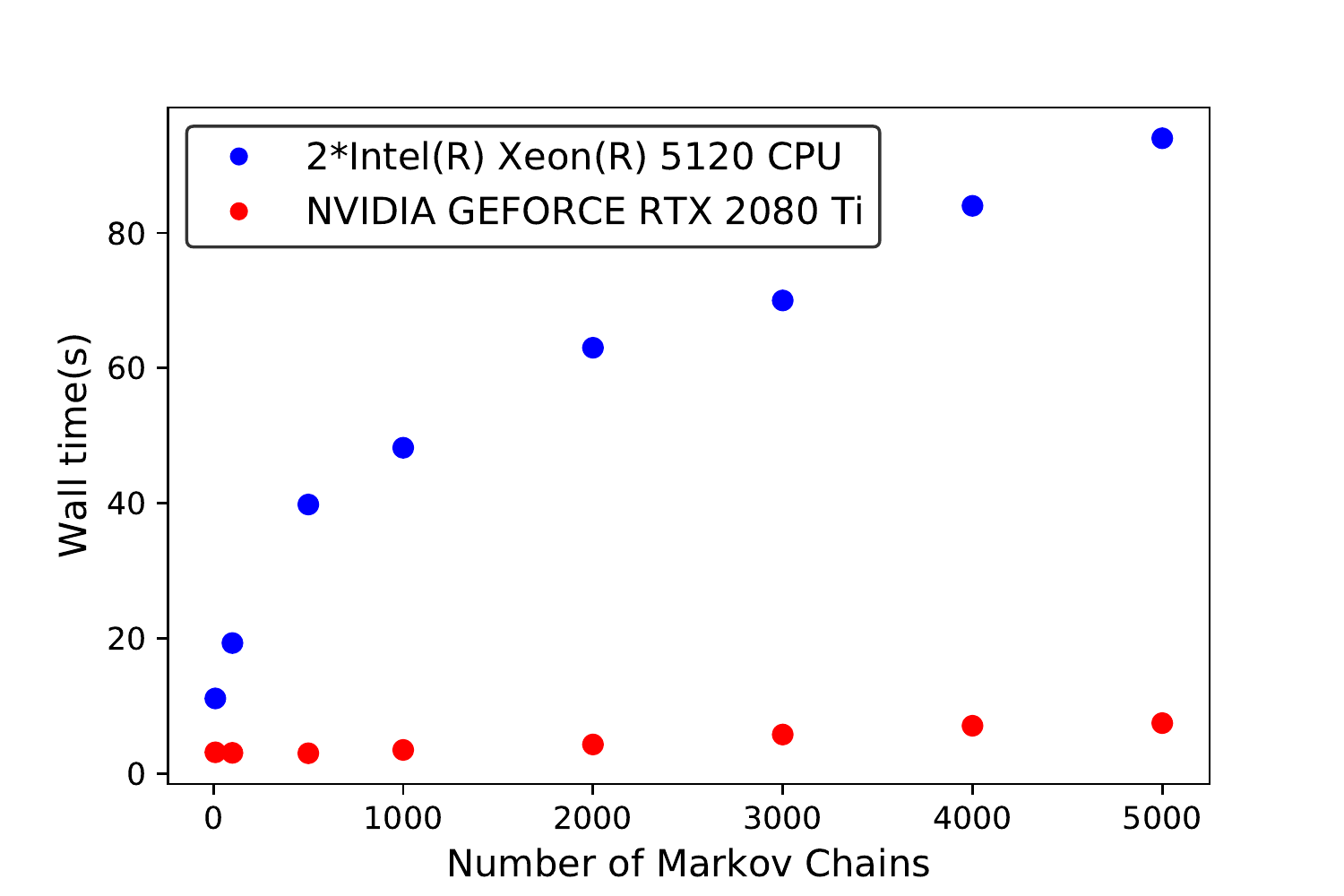}
	\includegraphics[width=0.35\textwidth]{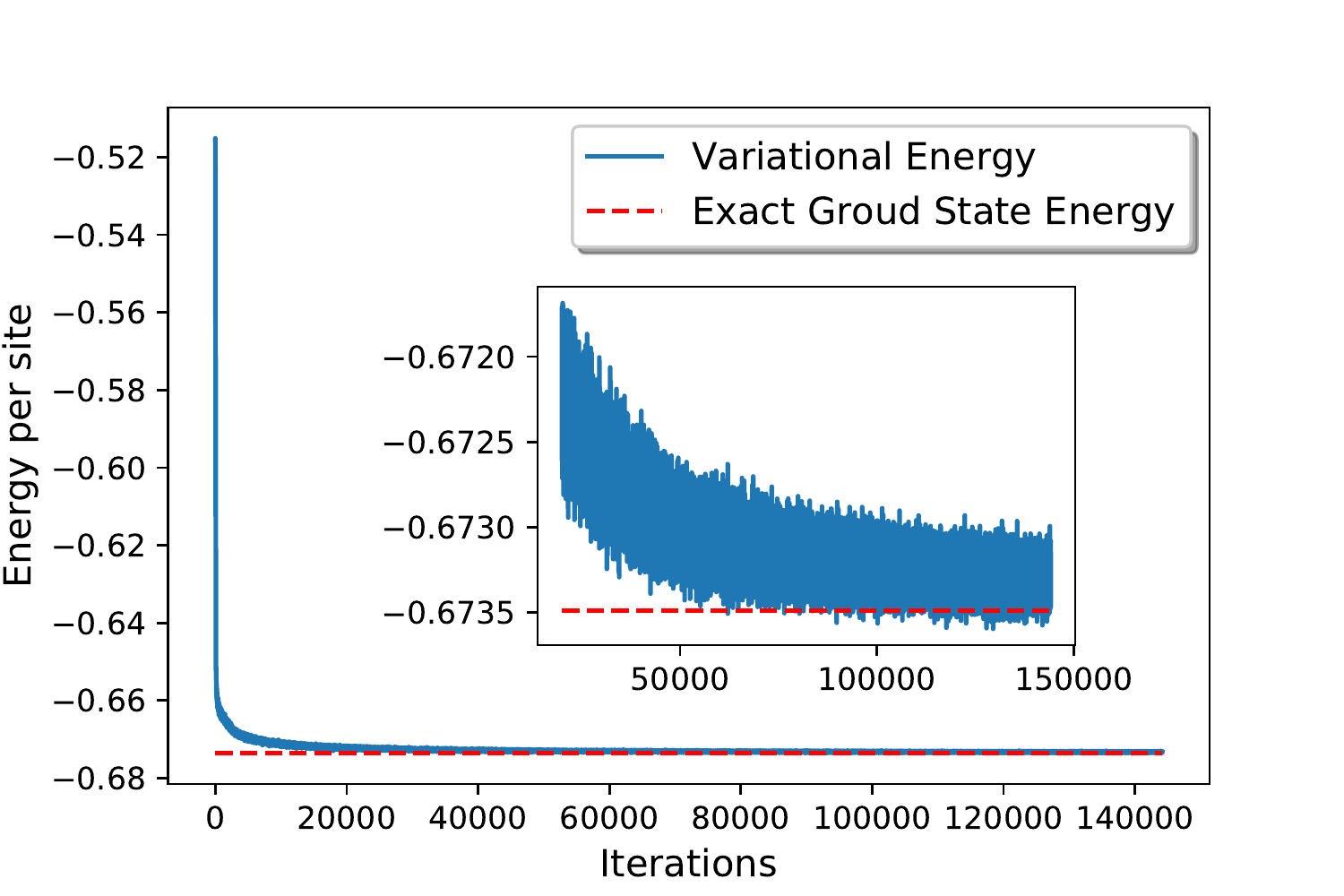}
	\caption{Numerical results of end-to-end ADVMC approach for the spin-1/2 quantum Heisenberg model on the square lattice. (a) Comparison of wall time on GPU and CPU by measuring the wall time of 20 iterations with different numbers of Markov chains. It was conducted with the same neural network states and model setup. (b) ADVMC results on $8\times 8$ quantum Heisenberg model with periodic boundary condition. The variational wave function was chosen to be a fully connected neural network with 7 layers. The number of nodes on each layer is $(16l^2, 8l^2,4l^2,8l^2,4l^2,l^2,1)$, with $l=8$. The activations were set to be RELU for all these layers except the last one. Adam optimizer was used to update the parameters. The dash line is the benchmark ground state energy given by the SSE method. Inset shows the converge of energy near the exact value. The ADVMC result is energetically competitive or advantageous over resulted obtained by various state-of-the-art methods including EPS, PEPS and RBM-based VMC.}
	\label{fig:subfig}
\end{figure}

Here we demonstrate this new paradigm of VMC by ADVMC study of the spin-1/2 quantum Heisenberg model on the square lattice. The model is given by $H=\sum_{\langle ij\rangle} J\vec S_i\cdot \vec S_j$, where $\vec S_j$ is spin-1/2 operator on site $j$. Because of the Marshal-sign rule the ground wave function amplitudes of the Heisenberg model can be rendered positive definite. Consequently, for simplicity we shall use positive ansatz wave functions in our ADVMC simulations. The computational graph we utilize in this problem is a fully-connected neural network with 7 layers and with RELU activations \cite{Hinton}. The number of nodes on these layers are $16l^2, 8l^2,4l^2,8l^2,4l^2,l^2,1$, where $l^2=64$ is the size of the system. Such neural network design is general without considering any symmetry and geometric information. Totally there are more than one million variational parameters and the number of Markov chains is 5000 in our ADVMC simulations. With such large amount of independent Markov chains and variational parameters, the ADVMC implementation is still very efficient on GPU in terms of time and storage resources due to the highly parallelized structure of our algorithm,  as shown in \Fig{fig:subfig}(a). The much shorter running time compared with CPU also demonstrates the increasing significance of GPU acceleration when the number of Markov chains increases.

The approximation ground state energy optimized by Adam converges to $-0.6733$ (in unit of $J$) per site averaged by the last 5000 energy data, as shown in \Fig{fig:subfig}(b). This result has $3\times10^{-4}$ relative error compared with the benchmark ground state energy obtained by SSE \cite{Sandvik1997}. It's also energetically competitive or advantageous over results obtained by various state-of-the-art methods including EPS \cite{Mezzacapo2009}, PEPS \cite{Wang2011, Lubasch2014} and RBM-based VMC \cite{Carleo2017}. This convincingly demonstrates that end-to-end ADVMC can enable us to reach state-of-the-art numerical results with very moderate effort for quantum models.

\section{Discussions and Conclusions} \label{sec:discussions}

One central issue in Monte Carlo simulations of interacting many-body quantum models is the notorious sign problem. Although it is shown to be NP hard to  solve the sign problem generically \cite{Troyer2005}, it is still possible to ease \cite{Hangleiter2019,Levy2019,Kim2019} or solve \cite{Li2015b} the sign problem of a given specific quantum model in QMC simulations by certain basis transformations. We propose to employ ADMC as a general way to find the optimal basis which can ease or solve the notorious sign problem in QMC simulations of interesting quantum models, such as the repulsive Hubbard model away half filling. One appropriate objective in ADMC would be the expectation value of the sign which depends on the parameters characterizing the basis choice. ADMC can help to find an optimal basis in which the sign problem is alleviated. 
From the ADMC-optimized basis with eased sign problem, one may simulate strongly correlated models with lower temperature and larger system size than QMC with usual basis.

ADMC proposed in the present paper is based on score function estimators. For the specific models we have studied, the present ADMC obtains accurate results without suffering any high variance in MC estimations. It is possible to further improve ADMC by reducing variance in MC estimations of expectations. In other words, it would be desired to find baselines or general control variables which could systematically reduce the variance of MC estimations. It is one of future routes to improve ADMC by introducing baselines suitable for any order derivatives as in the case of normalized probability distribution \cite{Mao2019}.

In conclusion, we have presented the general theory and framework of ADMC. We also showed how Monte Carlo expectations, KL divergence, and objectives from various settings can be expressed in an infinitely AD fashion. We further applied the ADMC approach on various Monte Carlo applications including classical Monte Carlo and end-to-end VMC. Especially, the ADVMC enables us to efficiently study interacting quantum models in higher dimensions. We believe that the ADMC approach can inspire more accurate and efficient Monte Carlo designs with machine learning toolbox in the future. At the intersection of differentiable programming and probabilistic programming, ADMC framework provides a promising route to advance Monte Carlo applications in the fields of statistics, machine learning, and physics. 
\newline

{\it Acknowledgement}:
We thank Shuai Chen, Zi-Xiang Li, Jin-Guo Liu,  Rong-Yang Sun and Lei Wang for helpful discussions. This work is supported in part by the NSFC under Grant No. 11825404 (SXZ, ZQW, and HY), the MOSTC under Grant Nos. 2016YFA0301001 and 2018YFA0305604 (HY), and the Strategic Priority Research Program of Chinese Academy of Sciences under Grant No. XDB28000000 (HY). HY would also like to acknowledge support in part by the Gordon and Betty Moore Foundations EPiQS Initiative through Grant GBMF4302 at Stanford.

\begin{widetext}
	\section*{Supplemental Materials}
	\renewcommand{\theequation}{A\arabic{equation}}
	\setcounter{equation}{0}
	\renewcommand{\thefigure}{A\arabic{figure}}
	\setcounter{figure}{0}
	
	\subsection{Automatic differentiation approach for Fisher information matrix}
	In this part, we further discuss the implementation details and advantages on AD approach towards FIM.
	
	The test case for algorithm implementations of FIM we utilized is simple distributions such as multivariate Gaussian distribution $N(\vect{\mu}\vert\vect{\sigma})$, in which $\vect{\mu}, \vect{\sigma}$ depend on variational parameters $\vect{\theta}$. For the simplest case, $\vect{\sigma}$ is constant and $\vect{\mu}(\vect{\theta})$ is determined by parameters $\vect{\theta}$. We can obtain analytical expression for FIM in this case:
	
	\eq{F_{ij}=\frac{\p \vect{\mu}^T}{\p \theta_i}\vect{\sigma}^{-1}\frac{\p \vect{\mu}}{\p \theta_j}.}{arg2}
	If we further assume $\vect{\sigma}=I$ and $\mu_i=\mu(\theta_i)$ is in the same function form, we can further simplify FIM analytically as:
	\eq{F=(\p \mu)^2I.}{arg2}
	
	In our code example, we test with three dimensional Gaussian distribution and $\mu(\theta)=(\theta+1)^2$ where $\theta = 0$. The expected FIM should be $4\,I_{3\times 3}$ in this case. Such test cases can also be used for testing implementation of unnormalized probability cases if the normalization factor of Gaussian distribution is deliberately dropped out.
	
	The first advantage for FIM with AD approach is zero elements might be kept without MC fluctuations or error bars. Take the test case above for an example, all off diagonal elements of FIM should be zero analytically. If one utilized conventional way computing FIM by MC averaging first order derivatives of $\ln p$, the resulting off diagonal elements are not zero due to the error bar introduced by MC. However, with advanced graph optimization and smart compiler infrastructure provided by TensorFlow, unnecessary computations can be identified and removed from runtime graph. With such state-of-the-art executing engine of computational graph, the off diagonal terms can be pinned at zero with AD approach. This is because the zero nature of these terms have already been identified at graph building time by TensorFlow engines. That is to say, the numerical result can even reach theoretical precision with the help of AD. It is worth noting that such gain is not guaranteed since TensorFlow engine can fail recognizing complicated series of unnecessary operations. For example, AD with unnormalized probability objectives give nonzero off diagonal elements in FIM using the same Gaussian distribution test case.
	
	The second advantage of AD approach is the better compatibility with vectorization scheme. Suppose we vectorize Markov chains as the batch dimension as the case in our implementation of example applications. The conventional way to evaluate FIM involving terms like $ \sum_{x\sim p}\p_i \ln p(x)\, \p_j \ln p(x)$ where $x$ is different configurations living on the extra vectorization dimension in our setup. It would be hard to evaluate such terms by treating the batch dimension as a whole where $x$ are different for different chains. This restriction is mainly brought by modern AD infrastructure of ML libraries in which derivatives for multiple outputs can only be obtained one by one and no tensorized fashion AD is implemented. Instead, KL divergence objective only concern about terms like $\sum_{x\sim p} \frac{p}{\bot p}$ which is super easy to parallelize by a simple reduce mean. The computation time of the conventional approach is scaling with the number of Markov chains or configuration samples $N$ which is typical thousands to millions while the computation time of AD approach is scaling with the parameter number (one has to apply AD on each derivatives to get the Hessian in ML libraries) which could be way less than the configuration numbers vectorized in the batch dimension. And our numerical experiments indeed show that AD approach is clearly faster than conventional approach either in graph building time or in graph executing time.

	\subsection{End-to-end ADVMC setup for general complex wave functions}
	
	\subsubsection{Computational graph setup for general wave function}
	
	\noindent If the ground state wave function is not always real positives, the general form can be expressed as $\psi_{\sigma}= e^{r_{\sigma}}e^{i\theta_{\sigma}}$, where $r$ characterize the real norm part $\ln\abs{\psi}$ and $\theta$ characterize the complex angle for the wave function. Therefore, we need two separate computational graphs for computing $r$ and $\theta$, and train them together towards minimal energy. We discuss about the most reliable form of AD-aware energy estimators and the assistant estimator for natural gradients in the following.
	
	\subsubsection{Infinite order AD estimator for VMC}
	
	The reason why VMC works is due to the following fact: the quantum expectation energy can be approximated by classical Monte Carlo averaged $E_{\mathrm{loc}}$.
	\eq{
		\begin{split}
			\left <H\right >&=\frac{\sum_{\sigma\sigma'}\psi^*_\sigma H_{\sigma\sigma'}\psi_{\sigma'}}{\sum_{\sigma}\psi_\sigma^*\psi_\sigma}=\frac{\sum_{\sigma\sigma'}\psi^*_\sigma\psi_\sigma (H_{\sigma\sigma'}\psi_{\sigma'}/\psi_\sigma)}{\sum_{\sigma}\psi_\sigma^*\psi_\sigma}=\frac{\sum_{\sigma}\psi^*_\sigma\psi_\sigma \sum_{\sigma'}(H_{\sigma\sigma'}\psi_{\sigma'}/\psi_\sigma)}{\sum_{\sigma}\psi_\sigma^*\psi_\sigma}\\
			&=\frac{\sum_{\sigma}\psi^*_\sigma\psi_\sigma E_{\mathrm{loc}}(\sigma)}{\sum_{\sigma}\psi_\sigma^*\psi_\sigma}=\Re\frac{\sum_{\sigma}\psi^*_\sigma\psi_\sigma E_{\mathrm{loc}}(\sigma)}{\sum_{\sigma}\psi_\sigma^*\psi_\sigma}=\frac{\sum_{\sigma}\psi^*_\sigma\psi_\sigma \Re E_{\mathrm{loc}}(\sigma)}{\sum_{\sigma}\psi_\sigma^*\psi_\sigma},\\
		\end{split}
	}{}
	where $E_{\mathrm{loc}}(\sigma)=\sum_{\sigma'}(H_{\sigma\sigma'}\psi_{\sigma'}/\psi_\sigma)$, the summation over $\sigma'$ can be done efficiently because $H_{\sigma\sigma'}$ is sparse matrix for general local Hamiltonian. If we treat $\psi_\sigma^*\psi_\sigma$ as the classical probability $p(\sigma)$, then we have
	\eq{\left< H\right>=\sum_{\sigma}{p(\sigma)(\Re E_{\mathrm{loc}}(\sigma))}/\sum_\sigma p(\sigma)=\left<\Re E_{\mathrm{loc}}(\sigma)\right>_{\sigma\sim p(\sigma)},}{}
	which indicates $\left< H\right>$ is just the expected value of $\Re E_{\mathrm{loc}}(\sigma)$ when $\sigma$ is sampled from an unnormalized distribution $p(\sigma)$. For this problem, our ADMC approach gives an accurate infinite order AD-aware estimator of it:
	\eq{
		\mathbf{Es} \left< H\right>= \frac{\sum_{\sigma\in S(p)}\frac{p_\sigma}{\bot p_\sigma}\Re E_{\mathrm{loc}}(\sigma)}{\sum_{\sigma\in S(p)}\frac{p_\sigma}{\bot p_\sigma}}=\frac{\sum_{\sigma\in S(p)}\frac{\psi_\sigma^*\psi_\sigma}{\bot (\psi_\sigma^*\psi_\sigma)}\Re E_{\mathrm{loc}}(\sigma)}{\sum_{\sigma\in S(p)}\frac{\psi_\sigma^*\psi_\sigma}{\bot (\psi_\sigma^*\psi_\sigma)}}.
	}{vmc_gest}
	Here $\sum_{\sigma\in S(p)}$ means doing summation on the set of configurations $\sigma$ sampled from distribution $p=\psi^*_{\sigma}\psi_{\sigma}$ using MCMC method. This estimator is correct for arbitrary order derivatives no matter whether the wave function ansatz is real or not. Nevertheless, we can design more efficient estimators in VMC context with lower variance and better optimization results as we shown below.

	\subsubsection{First order efficient AD estimator for VMC}
	
	In most of the cases, the knowledge about the gradients (first order derivatives) of $\left<H\right>$ is enough, while our estimator in \Eq{vmc_gest} is correct for any order of derivatives. There's possibility that we can further increase our precision if we focus on the first order derivative.
	
	By analytically deriving the gradients of $\left<H\right>$ from quantum expectation perspective, we have:
	\eq{
		\begin{split}
			\nabla \left<H\right>=&\nabla \frac{\sum_{\sigma\sigma'}\psi^*_\sigma H_{\sigma\sigma'}\psi_{\sigma'}}{\sum_{\sigma}\psi_\sigma^*\psi_\sigma}\\
			=&\frac{\sum_{\sigma\sigma'}(\nabla \psi^*_\sigma) H_{\sigma\sigma'}\psi_{\sigma'}+\psi^*_\sigma H_{\sigma\sigma'}(\nabla\psi_{\sigma'})}{\sum_\sigma{\psi_\sigma^*\psi_\sigma}}-\frac{\sum_{\sigma\sigma'}\psi^*_\sigma H_{\sigma\sigma'}\psi_{\sigma'}}{\sum_{\sigma}\psi_\sigma^*\psi_\sigma}\frac{\nabla \sum_{\sigma}\psi_\sigma^*\psi_\sigma}{\sum_{\sigma}\psi_\sigma^*\psi_\sigma}\\
			\overset{H_{\sigma\sigma'}=H^*_{\sigma'\sigma}}{\implies}&\frac{\sum_{\sigma}(\psi_\sigma^*\psi_\sigma)(\frac{\nabla \psi^*_\sigma}{\psi^*_\sigma} E_{\mathrm{loc}}(\sigma)+\frac{\nabla \psi_\sigma}{\psi_\sigma} E^*_{\mathrm{loc}}(\sigma))}{\sum_\sigma{\psi_\sigma^*\psi_\sigma}}-\frac{\sum_{\sigma\sigma'}\psi^*_\sigma H_{\sigma\sigma'}\psi_{\sigma'}}{\sum_{\sigma}\psi_\sigma^*\psi_\sigma}\frac{\sum_{\sigma}(\psi_\sigma^*\psi_\sigma)(\frac{\nabla \psi^*_\sigma}{\psi^*_\sigma}+\frac{\nabla \psi_\sigma}{\psi_\sigma})}{\sum_{\sigma}\psi_\sigma^*\psi_\sigma}\\
			=&2\Re(\frac{\sum_{\sigma}\psi_\sigma^*\psi_\sigma\frac{\nabla \psi^*_\sigma}{\psi^*_\sigma} E_{\mathrm{loc}}(\sigma)}{\sum_\sigma{\psi_\sigma^*\psi_\sigma}}-\frac{\sum_{\sigma}\psi^*_\sigma\psi_\sigma E_{\mathrm{loc}}(\sigma)}{\sum_{\sigma}\psi_\sigma^*\psi_\sigma}\frac{\sum_{\sigma}\psi_\sigma^*\psi_\sigma\frac{\nabla \psi^*_\sigma}{\psi^*_\sigma}}{\sum_{\sigma}\psi_\sigma^*\psi_\sigma}).
		\end{split}
	}{}
	
	Note all the terms are in the form of $\frac{\sum_\sigma \psi_\sigma^*\psi_\sigma O(\sigma)}{\sum_\sigma \psi_\sigma^*\psi_\sigma}$, as this kind of terms can be estimated by $\sum_{\sigma\in S(p)}O(\sigma)/N_{mc}$. Thus we have:
	\eq{
		\nabla \left<H\right>\doteq 2\Re(\frac{\sum_{\sigma\in S(p)}\frac{\nabla \psi^*_\sigma}{\psi^*_\sigma} E_{\mathrm{loc}}(\sigma)}{N_{mc}}-\frac{\sum_{\sigma\in S(p)} E_{\mathrm{loc}}(\sigma)}{N_{mc}}\frac{\sum_{\sigma\in S(p)}\frac{\nabla \psi^*_\sigma}{\psi^*_\sigma}}{N_{mc}}).
	}{vmc1}
	
	From MC expectation perspective, we calculate the gradients of the general estimator in \Eq{vmc_gest}, the result is not the same as \Eq{vmc1}, The difference terms are:
	\eq{
		\begin{split}
			\mathbf{diff}&=\frac{1}{N_{mc}}\Re\sum_{\sigma\in S(p)}(\frac{\nabla \psi^*_\sigma}{\psi^*_\sigma} E_{\mathrm{loc}} -\frac{\nabla \psi_\sigma}{\psi_\sigma} E_{\mathrm{loc}}-\nabla {E_{\mathrm{loc}}})\\
			&=\frac{1}{N_{mc}}\Re\sum_{\sigma\in S(p)}(\frac{\nabla \psi^*_\sigma}{\psi^*_\sigma} E_{\mathrm{loc}} -\frac{\nabla \psi_\sigma}{\psi_\sigma} E_{\mathrm{loc}}-\nabla \sum_{\sigma'}H_{\sigma\sigma'}\frac{\psi_{\sigma'}}{\psi_\sigma})\\
			&=\frac{1}{N_{mc}}\Re\sum_{\sigma\in S(p)}(\frac{\nabla \psi^*_\sigma}{\psi^*_\sigma} E_{\mathrm{loc}}-\sum_{\sigma'}H_{\sigma\sigma'}\frac{\nabla \psi_{\sigma'}}{\psi_\sigma}).
		\end{split}
	}{}
	$\mathbf{diff}$ normally is not zero numerically, but it goes to zero when $N_{mc}$ goes to infinite as it should be. This is because:
	\eq{
		\begin{split}
			\frac{1}{N_{mc}}\Re\sum_{\sigma\in S(p)}\sum_{\sigma'}H_{\sigma\sigma'}\frac{\nabla \psi_{\sigma'}}{\psi_\sigma}\overset{N_{mc}\rightarrow \infty}{=}&\Re\sum_{\sigma\sigma'}\psi^*_\sigma\psi_\sigma H_{\sigma\sigma'}\frac{\nabla \psi_{\sigma'}}{\psi_\sigma}/\sum_{\sigma}\psi^*_\sigma\psi_\sigma\\
			=&\Re\sum_{\sigma\sigma'}\psi^*_{\sigma'}\psi_{\sigma'}\frac{\nabla \psi_{\sigma'}}{\psi_{\sigma'} }H_{\sigma\sigma'}\frac{\psi^*_\sigma}{\psi^*_{\sigma'}}/\sum_{{\sigma'}}\psi^*_{\sigma'}\psi_{\sigma'}\\
			\overset{H_{\sigma\sigma'}=H^*_{\sigma'\sigma}}{=}&\Re\sum_{\sigma'}\psi^*_{\sigma'}\psi_{\sigma'}\frac{\nabla \psi_{\sigma'}}{\psi_{\sigma'} }\sum_{\sigma}(H_{\sigma'\sigma}\frac{\psi_\sigma}{\psi_{\sigma'}})^*/\sum_{{\sigma'}}\psi^*_{\sigma'}\psi_{\sigma'}\\
			=&\Re\sum_{\sigma'}\psi^*_{\sigma'}\psi_{\sigma'}\frac{\nabla \psi_{\sigma'}}{\psi_{\sigma'} }E^*_{\mathrm{loc}}(\sigma')/\sum_{{\sigma'}}\psi^*_{\sigma'}\psi_{\sigma'}\\
			=&\frac{1}{N_{mc}}\Re\sum_{\sigma\in S(p)}\frac{\nabla \psi^*_\sigma}{\psi^*_\sigma} E_{\mathrm{loc}}.
		\end{split}
	}{}
	Thus we proved $\lim_{N_{mc}\rightarrow \infty}\mathbf{diff}= 0$. In other words, If we directly use \Eq{vmc_gest} as the estimator, AD will give the gradients with another term whose expectation is theoretically zero.  Since this term is in general nonzero in MC calculations, it add more variance to the estimation on energy. We can safely drop $\mathbf{diff}$ term from the philosophy of baseline method. In other words, it would be better to find the estimator whose first order derivative is directly \Eq{vmc1} without $\mathbf{diff}$ term.
	
	Such estimator is easy to construct.
	
	\eq{\mathbf{Es_1}\left< H\right>=2\Re \frac{\sum_{\sigma\in S(p)}\frac{\psi^*_\sigma}{\bot\psi^*_\sigma}\bot E_{\mathrm{loc}}(\sigma)}{\sum_{\sigma\in S(p)}\frac{\psi^*_\sigma}{\bot\psi^*_\sigma}}.}{first-order-vmc}
	We can prove it by directly calculating the gradient of this estimator:
	
	\eq{\nabla \mathbf{Es_1\left<H\right>}=2\Re \frac{\sum_{\sigma\in S(p)}\frac{\nabla \psi^*_\sigma}{\bot\psi^*_\sigma}\bot E_{\mathrm{loc}}(\sigma)\sum_{\sigma\in S(p)}\frac{\psi^*_\sigma}{\bot\psi^*_\sigma}-\sum_{\sigma\in S(p)}\frac{\psi^*_\sigma}{\bot\psi^*_\sigma}\bot E_{\mathrm{loc}}(\sigma)\sum_{\sigma\in S(p)}\frac{\nabla \psi^*_\sigma}{\bot\psi^*_\sigma}}{(\sum_{\sigma\in S(p)}\frac{\psi^*_\sigma}{\bot\psi^*_\sigma})^2}.}{}
	
	In the sense of its numerical value
	
	\eq{\bot \nabla \mathbf{Es_1\left<H\right>}=\bot 2\Re(\frac{\sum_{\sigma\in S(p)}\frac{\nabla\psi_\sigma^*}{\psi_\sigma^*}E_{\mathrm{loc}}(\sigma)}{N_{mc}}-\frac{\sum_{\sigma\in S(p)} E_{\mathrm{loc}}(\sigma)}{N_{mc}}\frac{\sum_{\sigma\in S(p)}\frac{\nabla \psi^*_\sigma}{\psi^*_\sigma}}{N_{mc}}),}{}
	which is just the same as \Eq{vmc1}. Thus AD-aware estimator in \Eq{first-order-vmc} gives the right approximation of the gradients of $\left<H\right>$ with lower variance than the general estimator \Eq{vmc_gest}. Though it is only valid for the first order derivatives.
	
	If the wave function is real, \Eq{first-order-vmc} reduce to
	\eq{\mathbf{Es}_{1r}=\frac{\sum_{\sigma\in S(p)}\frac{p_\sigma}{\bot p_\sigma}\Re \bot E_{\mathrm{loc}}(\sigma)}{\sum_{\sigma\in S(p)}\frac{p_\sigma}{\bot p_\sigma}}.}{realfirst}
	One can again verify it by directly differentiating on \Eq{realfirst}. The only change in \Eq{realfirst} is detached $E_{\mathrm{loc}}$  comparing with the general estimator \Eq{vmc_gest}. The new estimator also has better performance when running the computational graph since $E_{\mathrm{loc}}$ is detached and no backward propagation pass through it.
	
	\subsubsection{SR and Natural gradients}
	
	SR method (natural gradient descent) is reported to give faster convergence on VMC. In this part, we explore the relation between natural gradient descent and SR method in general settings where the wave function could be complex.
	
	For real wave function case, KL divergence plays the role as the distance of distribution space whose Hessian FIM gives the same formalism as SR method as we have shown in the main text. In terms of complex case, traditional KL divergnece defined with $p=\psi^*\psi$ loses the information of wave function's phases. Thus we need a better distance measure to describe the difference between different wave functions.
	
	The natural choice is Fubini-Study distance defined in Hilbert space:
	
	\eq{s(\psi,\phi)= \arccos\sqrt{\frac{\left<\psi|\phi\right>\left<\phi|\psi\right>}{\left<\psi|\psi\right>\left<\phi|\phi\right>}}.}{fs}
	
	Infinitesimal distances are thus given by:
	
	\eq{ ds^2=s(\psi,\psi+\delta\psi)^2= \frac{\left<\delta\psi|\delta\psi\right>}{\left<\psi|\psi\right>} -\frac{\left<\delta\psi|\psi\right>}{\left<\psi|\psi\right>}\frac{\left<\psi|\delta\psi\right>}{\left<\psi|\psi\right>} =\left< \frac{\delta\psi_\sigma}{\psi_\sigma}\frac{\delta \psi_\sigma^*}{\psi_\sigma^*}
		\right>_{\sigma\sim \psi^*\psi}-\left< \frac{\delta\psi_\sigma}{\psi_\sigma}
		\right>_{\sigma\sim \psi^*\psi}\left< \frac{\delta \psi_\sigma^*}{\psi_\sigma^*}
		\right>_{\sigma\sim \psi^*\psi},
	}{FS-dis}
	where $\delta \psi =\p_i \psi d\theta_i$.
	
	Thus
	\eq{
		\begin{split}
			ds^2=&\sum_{\alpha,\beta}(\left< \frac{\partial_\alpha\psi_\sigma}{\psi_\sigma}\frac{\partial_\beta \psi_\sigma^*}{\psi_\sigma^*}
			\right>_{\sigma\sim \psi^*\psi}-\left< \frac{\partial_\alpha\psi_\sigma}{\psi_\sigma}
			\right>_{\sigma\sim \psi^*\psi}\left< \frac{\partial_\beta \psi_\sigma^*}{\psi_\sigma^*}
			\right>_{\sigma\sim \psi^*\psi})d\theta_\alpha d\theta_\beta\\
			=&\sum_{\alpha,\beta}\Re(\left< \frac{\partial_\alpha\psi_\sigma}{\psi_\sigma}\frac{\partial_\beta \psi_\sigma^*}{\psi_\sigma^*}
			\right>_{\sigma\sim \psi^*\psi}-\left< \frac{\partial_\alpha\psi_\sigma}{\psi_\sigma}
			\right>_{\sigma\sim \psi^*\psi}\left< \frac{\partial_\beta \psi_\sigma^*}{\psi_\sigma^*}
			\right>_{\sigma\sim \psi^*\psi})d\theta_\alpha d\theta_\beta\\
			=&\sum_{\alpha,\beta} S_{\alpha\beta}d\theta_\alpha d\theta_\beta ,
		\end{split}
	}{}
	where
	\eq{
		S_{\alpha\beta}=\Re\left(\left< \frac{\partial_\alpha\psi_\sigma}{\psi_\sigma}\frac{\partial_\beta \psi_\sigma^*}{\psi_\sigma^*}
		\right>_{\sigma\sim \psi^*\psi}-\left< \frac{\partial_\alpha\psi_\sigma}{\psi_\sigma}
		\right>_{\sigma\sim \psi^*\psi}\left< \frac{\partial_\beta \psi_\sigma^*}{\psi_\sigma^*}
		\right>_{\sigma\sim \psi^*\psi}\right)
	}{FIM}
	is identical to quantum version of Fisher Information Matrix utilized in SR method.
	
	\Eq{FIM} can be estimated by ADMC approach, that is:
	\eq{
		\begin{split}
			S_{\alpha\beta}&\doteq\Re(\left< \frac{\partial_\alpha\psi_\sigma}{\psi_\sigma}\frac{\partial_\beta \psi_\sigma^*}{\psi_\sigma^*}
			\right>_{\sigma\in S(p)}-\left< \frac{\partial_\alpha\psi_\sigma}{\psi_\sigma}
			\right>_{\sigma\in S(p)}\left< \frac{\partial_\beta \psi_\sigma^*}{\psi_\sigma^*}
			\right>_{\sigma\in S(p)})\\
			&=\Re(\left< \partial_\alpha\ln\psi_\sigma\partial_\beta \ln\psi_\sigma^*
			\right>_{\sigma\in S(p)}-\left< \partial_\alpha\ln\psi_\sigma
			\right>_{\sigma\in S(p)}\left< \partial_\beta \ln \psi_\sigma^*
			\right>_{\sigma\in S(p)})\\
			&=\left< \partial_\alpha\ln|\psi_\sigma|\partial_\beta \ln|\psi_\sigma|
			\right>_{\sigma\in S(p)}-\left< \partial_\alpha\ln|\psi_\sigma|
			\right>_{\sigma\in S(p)}\left< \partial_\beta \ln |\psi_\sigma|
			\right>_{\sigma\in S(p)}+\left< \partial_\alpha\theta_\sigma\partial_\beta \theta_\sigma
			\right>_{\sigma\in S(p)}-\left< \partial_\alpha\theta_\sigma
			\right>_{\sigma\in S(p)}\left< \partial_\beta \theta_\sigma
			\right>_{\sigma\in S(p)}.
		\end{split}
	}{FIMMC}
	Using detach function, we also have the relationship already utilized in FIM formalisms with unnormalized distribution $p$:
	\eq{
		\begin{split}
			\bot\partial^2_{\alpha\beta}\left(\ln\left<\frac{O_\sigma}{\bot O_\sigma}\right>_{\sigma\in S(p)}-\left<\ln\frac{O_\sigma}{\bot O_\sigma}\right>_{\sigma\in S(p)}\right)
			&=\bot\partial_{\alpha}\left(\frac{\left<\frac{\partial_\beta O_\sigma}{\bot O_\sigma}\right>_{\sigma\in S(p)}}{\left<\frac{O_\sigma}{\bot O_\sigma}\right>_{\sigma\in S(p)}}-\left<\frac{\partial_\beta O_\sigma}{ O_\sigma}\right>_{\sigma\in S(p)}\right)\\
			&=\bot\left(\left<\partial_\alpha O_\sigma \partial_\beta O_\sigma \right>_{\sigma\in S(p)}-\left<\partial_\alpha O_\sigma  \right>_{\sigma\in S(p)}\left< \partial_\beta O_\sigma \right>_{\sigma\in S(p)}\right).
		\end{split}
	}{relation}
	Note detach function at the beginning is used to emphasize this relationship is only true in value for the second derivatives of LHS.
	
	Considering the general computational graph setup with two graphs $r$ and $\theta$ which gives $\psi_\sigma=e^{r_{\sigma}}e^{i\theta_\sigma}$. Using this relationship, \Eq{FIMMC} can be calculated as the Hessian of an AD-aware estimator. The estimator is:
	\eq{
		\mathbf{Es}_{ng}=\ln\left<\frac{r_{\sigma}}{\bot r_\sigma}\right>_{\sigma\in S(p)}-\left<\ln\frac{r_\sigma}{\bot r_\sigma}\right>_{\sigma\in S(p)}+\ln\left<\frac{\theta_\sigma}{\bot \theta_\sigma}\right>_{\sigma\in S(p)}-\left<\ln\frac{\theta_\sigma}{\bot \theta_\sigma}\right>_{\sigma\in S(p)}.
	}{Es_sr}
	This estimator can be viewed as the generalized version of KL divergence in complex distribution space.
	
	For real positive wave function case, there will be no $\theta$ term, the Hessian of $\mathbf{Es}_{ng}$ is just $\mathbf{FIM}/4$, where $\mathbf{FIM}$ is classical Fisher Information Matrix, i.e. the Hessian of conventional KL divergence.
	
	To summarize, the natural distance measure in wave function Hilbert space is Fubini-Study distance as in \Eq{fs}. The Hessian of it gives the inverse matrix to be applied before the gradients utilized in natural gradient method. From the implementation perspective, such distance can be substituted by the extended version of KL divergence as in \Eq{Es_sr}. In this context, the Hessian of extended KL estimator, quantum version of FIM, the Hessian of Fubini-Study distance and the matrix required in SR method are literally the same thing. All these objects are connecting with each other and it is interesting to see how SR method can emerge in the context of natural gradient descent from information geometry without knowledge about imaginary time evolution by Schr\"odinger equation.
	
	It is worth noting in SR method, we need to inverse FIM for natural gradient descent. However, FIM is often peculiar with very large condition number rendering the inverse of FIM numerically unstable. The singular spectrum of FIM has been investigated very recently \cite{Karakida2019a,Park2019}. From implementation perspective, the most simple workaround is adding $\varepsilon I$ on $F$ before inverse. 
	
\end{widetext}
\end{document}